\documentclass[pre,aps,amsmath,amssymb,superscriptaddress,notitlepage,twocolumn]{revtex4-2}
\usepackage[colorlinks,bookmarks=false,citecolor=blue,linkcolor=blue,urlcolor=blue]{hyperref}
\usepackage[all]{hypcap}   % let hyperlinks correctly point to figures rather than their captions;
\usepackage{soul}

\usepackage{placeins}
\usepackage{ulem} % or \usepackage[normalem]{ulem}
\usepackage{amsmath,amssymb,amssymb}
\usepackage{graphicx}
\graphicspath{{figures/}{/}}
\usepackage{dcolumn}
\usepackage{bm}
\usepackage[capitalise,compress]{cleveref}
\crefname{section}{Sec.}{Secs.}
\Crefname{section}{Section}{Sections}

\usepackage{verbatim}
\usepackage{color}
\usepackage{soul}
\usepackage{placeins}  % for FloatBarrier
\usepackage{flafter}     % Bilder I'm
\usepackage{color}
\usepackage{hhline}
\usepackage{physics}
\usepackage{hyperref}

\usepackage{float}
\usepackage{caption}
\usepackage{subcaption}
\usepackage{lipsum} % just for the example

\usepackage[mathscr]{eucal}
\usepackage{tikz}
\usepackage{xcolor}

\newcommand{\bo}{\begin{outline}}
\newcommand{\eo}{\end{outline}}

\newcommand{\qed}{\nobreak \ifvmode \relax \else
      \ifdim\lastskip<1.5em \hskip-\lastskip
      \hskip1.5em plus0em minus0.5em \fi \nobreak
      \vrule height0.75em width0.5em depth0.25em\fi}
\begin{document} 

%\title{Quantum resonance encryption for secure data storage and communication with quantum kicked top}
\title{Quantum resonance based encryption protocol with quantum kicked top}
\author{Sreeram PG}
\email{sreerampg7@gmail.com}
\affiliation{School of Computing, MIT Art, Design and Technology University, Pune 412201, India}
\affiliation{Department of Physics, Indian Institute of Science Education and Research, Pune 411008, India}

\author{M. S. Santhanam}
\email{santh@iiserpune.ac.in}
\affiliation{Department of Physics, Indian Institute of Science Education and Research, Pune 411008, India}

%\date{\today}

\begin{abstract}
%In a shared quantum computer, how to ensure data privacy and protection from access by unauthorized parties? We propose a genuine quantum protocol for protecting user's data which is not accessible even to the service provider. 
We propose a genuine quantum protocol for protecting user's data, either in a shared quantum computer or in a quantum communication system, that is not accessible even to the service provider. 
The protocol is based on quantum kicked top -- the dynamics of a spin system -- operating in the regime of quantum resonance. This protocol ensures perfect recovery for authorized users while making intercepted states appear mixed to eavesdroppers, with built-in tampering detection. This protocol can be used for secure communication between two parties in geographically different locations, and also for quantum key distribution. The effectiveness of this protocol is demonstrated by assuming a quantum computer with quantum memory and functioning quantum networks. In the absence of the latter, at present, the protocol can be demonstrated in a laboratory using currently available quantum computing platforms. 
 \end{abstract}
\maketitle
\section{Introduction} As quantum technology matures, fault-tolerant quantum machines with capability for end-to-end processing might become available. This also implies that attention must be paid to data privacy and security aspects, especially since sufficiently large fault-tolerant quantum computers are expected to crack current encryption standards \cite{mosca2018cybersecurity}. The classical public-key cryptography relies on a set of hard mathematical problems such as integer factorization, discrete logarithm, and elliptic curve discrete logarithm which classical computers cannot solve in polynomial time \cite{yan2013quantum}. However, they are vulnerable to Shor's algorithm \cite{shor1994algorithms} implemented on a large fault-tolerant quantum computer with as few as $10,000$ qubits \cite{cain2026shor}. This might be achievable in the near future. Against this backdrop, genuinely quantum data security protocols are required within the shared quantum computers as well as for secure communication between distant nodes. 

Existing quantum encryption protocols are primarily centered around quantum key distribution (QKD) protocols for generating and sharing secret keys between two parties that can then be used with classical encryption such as one-time pad or AES \cite{boykin2003optimal,daemen1999aes}. The central idea is that any eavesdropping attempt inevitably disturbs the quantum states and can be detected. More than three decades since the celebrated BB84 protocol \cite{bennett2014quantum} was introduced, many variants have been developed that  uses entanglement sharing \cite{farre2025entanglement} and   continuous quantum variables \cite{zhang2008secure}, other device-independent protocols that addresses detector side-channel vulnerabilities \cite{acin2007device,pironio2009device, vazirani2019fully, zhang2022device} and adaptive techniques that adapt to channel conditions in real time leading to improved key accuracy and security resilience. 

In recent years, quantum chaotic dynamics are increasingly employed for algorithmic applications, for scalable Grover search to efficient quantum reservoir computing \cite{romanelli2006quantum,ahmed2024prediction,kobayashi2026edge}. These approaches have the advantage that implementation of these algorithms does not require a gate-based quantum computer.
In the spirit of this approach, we propose an encryption protocol that combines quantum dynamics with quantum information principles, and provide a powerful layer of data security. The protocol works by obscuring the information to be protected by scrambling it over the large Hilbert space. The scrambling is quickly achieved by quantum chaotic dynamics of the well-studied the quantum kicked top \cite{haake1987classical,haake1991quantum, chaudhury2009quantum}. This has the advantage of experimental realization using an ensemble of spins periodically coupled to each other \cite{neill2016ergodic,krithika2019nmr}.  

In the proposed protocol, while quantum chaos ensures scrambling of information, its safe retrieval relies on quantum recurrences in the quantum kicked top \cite{anand2024quantum, zou2025enhancing}. This has no classical analog, and is termed quantum resonance. Thus, the core idea is to store or transmit the data in the scrambled state unintelligible to eavesdroppers. It can be later retrieved by completing the dynamical evolution for the recurrence period.  The parameter values of resonance, which are the key to retrieving the encrypted data, can be shared classically with someone else to grant them authorized access. 

We begin by describing the setup in Sec. \ref{sec: 2},  before discussing the quantum resonance encryption in Sec. \ref{sec: 3}. We follow it up by a quantum secure direct communication protocol and quantum key distribution using the same encryption, in sections \ref{sec: 4} and \ref{sec: 5} respectively. Practicability of the protocol in the full recurrence case is discussed in the end before concluding. 
\begin{figure}[!tbp]
    \centering
    \includegraphics[width=1\linewidth]{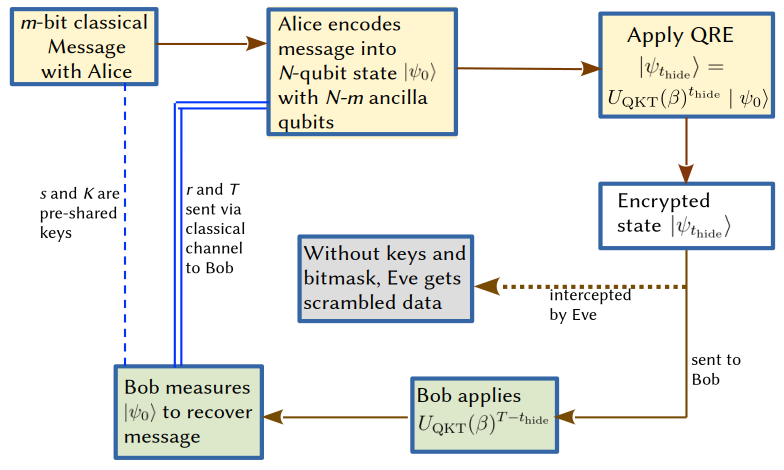}
    \caption{A schematic of the proposed protocol. Yellow shaded boxes are tasks done by Alice to prepare encrypted state $|\psi_{t_{\rm hide}}\rangle$. Light green shaded boxes are tasks done by Bob to recover the encrypted state. (grey shaded box) Eve can intercept the encrypted state during transmission, but will get only scrambled data upon measurement since she lacks the preshared keys as well as $r$ and $T$. Double line indicates data shared via classical channel, and dashed line indicates pre-shared parameters between Bob and Alice.}
    \label{fig: storage_access}
\end{figure}

\section{The protocol setup} \label{sec: 2}
The protocol consists of three parts; (a) data encoding into $N$-qubit state, (b) data encryption using quantum resonance, and (c) decryption using pre-shared keys and those inferred from quantum evolution at resonance. The protocol employs two sets of keys; a set that is pre-shared between Alice and Bob, and another set of keys that are obtained through time evolution of quantum kicked top at resonance. Figure \ref{fig: storage_access} shows the schematic of the protocol. 

\subsection{Data encoding}
Consider $N$ spin-$\frac{1}{2}$ particles (qubits) residing in Hilbert space $\mathcal{H}$ of dimension $d = 2^N$. Their collective spin quantum number is $j = N/2$. As usual, we consider two characters Alice and Bob who want to share secret information $\mathcal{I}$ accessible only to them and to no one else. Alice first converts $\mathcal{I}$ to binary form $\mathcal{I}_b$. She then prepares an $N$ qubit quantum register, in which $m$ (out of $N$) qubits are randomly chosen to encode $\mathcal{I}_b$. The message is encoded through mapping bits $0 \mapsto \ket{0}$ and $1 \mapsto \ket{1}.$ The remaining $N-m$ qubits are ancilla qubits initialized to $\ket{0}$. For instance, if $N=6,$ and given that the message qubits are represented by $\{\ket{s_i}\}, i=1,2,3$, the collective initial state could be
\begin{equation}
    \ket{\psi_0}=\ket{0} \otimes \ket{s_1} \otimes \ket{s_2} \otimes \ket{0} \otimes \ket{s_3} \otimes \ket{0},
\end{equation}
or any other permutations of the message and ancilla qubits. This provides a layer of security, as an eavesdropper (usually named Eve) who gains access to the physical qubits does not know which qubits  hold the message.

\subsection{Quatum kicked top}
The operational part of the algorithm -- encryption and decryption -- employ quantum kicked top (QKT), a well studied quantum chaotic system and hence we will only briefly describe it here.  The QKT is the dynamics of a spin vector $\vec{\mathbf J}$ on a unit sphere such the total squared spin angular momentum $\vec{\mathbf J}^2$ is conserved. The dynamics consists of a rotation about, say $y$-axis, followed by periodic application of position dependent rotation, { i.e.}, torsion, about $z$-axis. Due to periodicity of these operations, the quantum dynamics can be studied  through the Floquet theory \cite{shirley1965solution}. The period-one Floquet operator is
\begin{equation}
    U_{\text{QKT}} = \exp\left(-i \frac{\beta}{2j} J_y^2\right) \exp(-i \alpha J_z),
\end{equation}
where $J_x, J_y, J_z$ are collective angular momentum operators \cite{ghose2008chaos, dogra2019quantum,sreeram2021out}, $\alpha=\pi/2$ is a fixed  parameter, and $\beta$ is the kicking strength. The nonlinearity induced by the periodic torsion leads to quantum chaotic dynamics if strength of torsion $\beta$ is sufficiently large. Alice now couples the qubits through the kicked top Floquet evolution at resonance. 

\begin{figure}[t]
     \centering
     \includegraphics[width=0.9\linewidth]{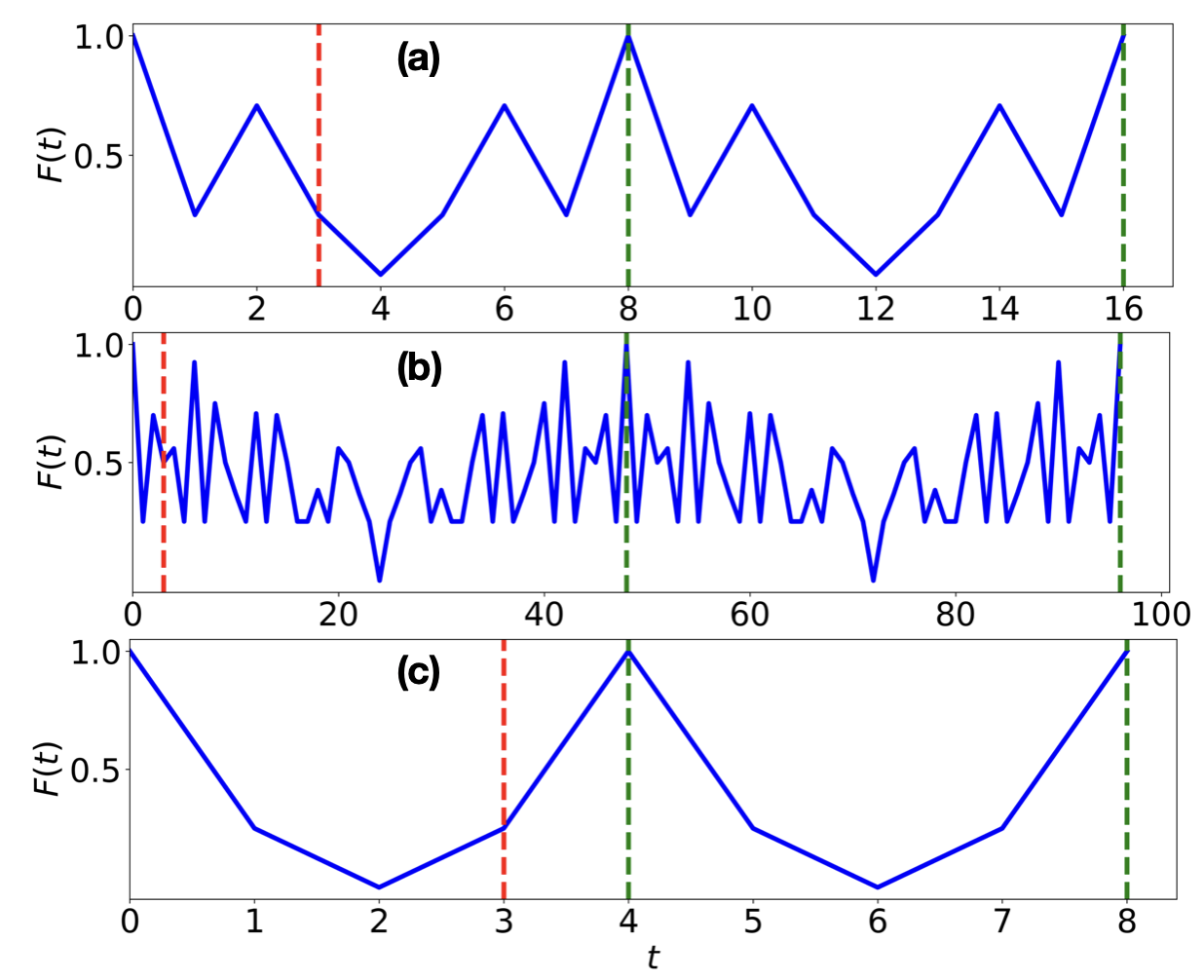}
     \caption{Fidelity recurrences of an { initial state} $\ket{1010}$ of a $j=2$ kicked top with pre-shared key $K=100$ and $r=1003,$ and $r_\mathrm{eff}=3$ for (a) $s=4$,  (b) $s=8$, and (c) $s=1$. Alice encodes the message with $B_\mathrm{decimal}=10,$  and  saves the encrypted state at ${t_\text{hide}}=3.$  The red dashed line marks $t_\mathrm{hide}$ and the green dashed lines mark the exact revival at $t=T$.}
    %Therefore, $r= 10 \times 100+3=1003.$
     \label{fig:exact_recorrence}
 \end{figure}
 
The encryption and decryption protocols make use of quantum resonance in QKT by choosing
\begin{equation}
    \beta = \frac{4\pi j r}{s} \label{beta}, \;\;\;\;\;\;\; r,s \in \mathbb{Z} ~\text{are co-primes.}
\end{equation}
For this choice of $\beta$, the quantum Floquet time-evolution of an initial state $\ket{\psi_0}$
brings it arbitrarily close to the initial state again after a time $T,$ giving $U^T\ket{\psi_0} \approx \ket{\psi_0}.$ However, there are special values of $\beta$ where exact recurrence occurs, where the Floquet time-evolution 
operator acts as an identity operator, $U^T = \mathbb{I}$, after a specific number of kicks $T$.
This phenomenon is called quantum resonance and is particularly interesting because classically at these parameter values the system is chaotic, and the quantum recurrence ignores the chaotic classical dynamics. Figure \ref{fig:exact_recorrence} displays exact recurrences of an initial state $|\psi_0\rangle$ through the fidelity $F(t) = |\langle \psi_0 | \psi(t) \rangle|^2$, which reaches 1 after period $T$. This figure shows two full periods marked by vertical dashed lines. Note that the pattern in each case (in Fig. \ref{fig:exact_recorrence}) precisely repeats itself at intervals of $T$. In Ref. \cite{anand2024quantum}, for kicked top with integer spin the observed periodicity of exact recurrence are $T=2,4,8,48$, while for half-integer spin top $T=4,12$. Since the protocols use these values of $T$, the exact recurrence offers few possibilities.

\begin{figure}[t]
     \centering
     \includegraphics[width=0.9\linewidth]{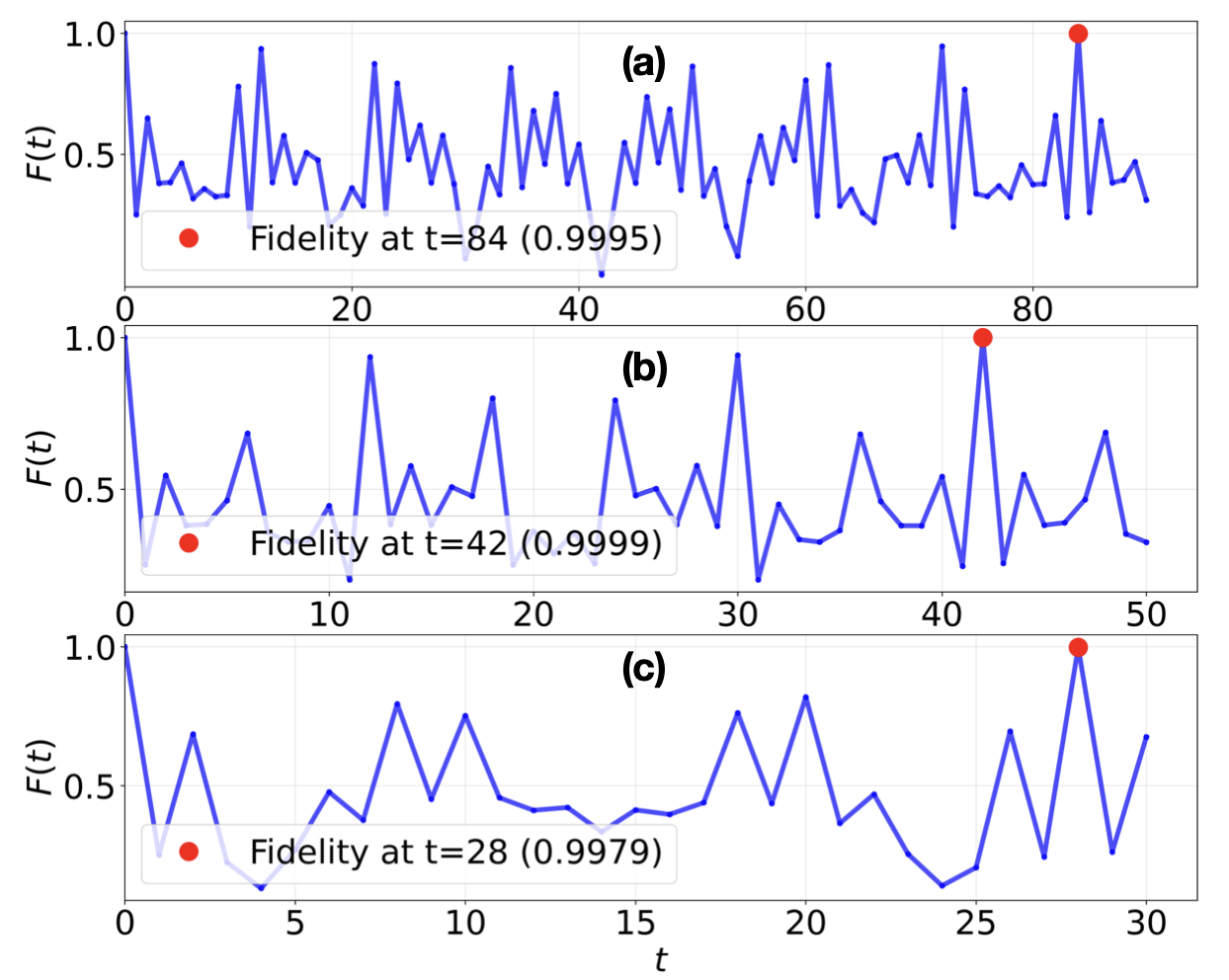}
     \caption{Near-recurrence of fidelity for a $j=2$ kicked top, with the initial state $\ket{1010}, B_\mathrm{decimal}=10, t_\mathrm{hide}=3.$ and pre-shared key $K=100$ for (a) $s=3$, (b) $s=6$, and (c) $s=7$. Hence, $r= B_\mathrm{decimal} \times K +t_\mathrm{hide}= 1003.$  The first time when $F(t) > 0.99$ is highlighted with a red circle. }
     \label{fig:almost_periodic}
 \end{figure}

However, if we relax the condition of exact recurrence, and instead consider \textit{near-recurrence} to an initial state, then many other possibilities for values of $T$ open up. To describe this situation, let the eigenvectors of the finite dimensional $U_\mathrm{QKT}$ be denoted by $\ket{k}$ and corresponding eigenvalues by $\exp(-i\phi_k),$ with $k=1,2, \dots 2j+1$, such that
\begin{equation}
    U_\mathrm{QKT}= \sum_{k=1}^{d=2j+1} \exp(-i\phi_k)\ket{k}\bra{k}.
\end{equation}
The initial state $\ket{\psi_0}$ in the Floquet eigenbasis is
\begin{equation}
    \ket{\psi_0}= \sum_k c_k\exp(-i\phi_k), \;\;\; \text{where} \;\;  c_k=\bra{k}\psi_0\rangle.
\end{equation}
Then the recurrence amplitude at time $t$ is
\begin{align}
    A_{\psi_0}(t) = \bra{\psi_0}U_\mathrm{QKT}^t\ket{\psi_0}
    = \sum_{k=1}^d |c_k|^2 \exp(-i\phi_kt).
\label{eq:recur_amp}
\end{align}
Note that the phases $\phi_k$ are finite in number. Then, the Kronecker's theorem, applied to Eq. \ref{eq:recur_amp}, guarantees that for any $\epsilon>0$, there exists a finite integer period $T$ such that all the phases are simultaneously within the $\epsilon$-neighbourhood of their initial values \cite{kronecker1884naherungsweise,bohr2018almost,gonek2016kronecker}. This allows a possibility that  
\begin{equation}
\bra{\psi_0}U_{\text{QKT}}^T\ket{\psi_0} \approx \mathbb{I}.     
\end{equation}
and consequently ensures the existence of { near-recurrence} for some integer period $T$. However, in  this case, it does not guarantee that subsequent near-recurrences occur with same period $T$. As the protocol requires only one near-recurrence to take place, the absence of subsequent recurrences will not affect the protocol. This near-recurrence can be loosely regarded as approximate resonance. The ``resonance'' timescale $T$ corresponding to the state $\ket{\psi_0}$ is defined as the first instant at which $A_{\psi_0}(t)>a$, where $a=0.99$. Figure \ref{fig:almost_periodic} demonstrates these near-recurrences for an initial state $\ket{1010},$ under three different $(r,s).$ In these cases, near-recurrence is identified once fidelity crosses 0.99. The actual fidelity values are indicated in Fig. \ref{fig:almost_periodic}. Note that in all the three case in this figure, $T$ is different and the next near-recurrence is not guaranteed to occur after time $T$.

\subsection{Quantum Resonance Encryption (QRE) } \label{sec: 3}
%\subsection{Encryption by Alice}
{First, we consider the exact resonance condition.}
The two parties who wish to exchange secure message will be called Alice and Bob. The message ($\mathcal{I}_b$) itself is encoded in an $N$-qubit initial state as
\begin{equation}
    \ket{\psi_0}=\ket{0} \otimes \ket{s_1} \otimes \ket{s_2} \otimes \ket{0} \dots \otimes \ket{0} \otimes \ket{s_m} \otimes \ket{0},
    \label{eq:initial_state}
\end{equation}
in which randomly chosen $N-m$ qubits hold the desired message. Thus, Eq. \ref{eq:initial_state} is just one possible realization for an initial state. 
As Bob is authorized to access the message from Alice, $s$ and $K>T$ are pre-shared keys (PSK) between Alice and Bob. Apart from PSK, there are dynamic keys generated by the encryption algorithm. The dynamic key is $r \in \mathbb{Z}$ that Alice computes for each encryption, which encodes both the message location and encryption timing. Alice generates an $N$-bit bitmask $B$ where bit $i$ is 1 if $i$-th qubit holds message, and $0$ if it is an ancillary qubit. Alice chooses  encryption time $t_{\rm hide}$ where $0 < t_{\rm hide} < T$.   Alice can choose $t_{\rm hide}$ such that the resultant state is not extremely fragile like a GHZ state, and can stay longer when stored, as the protocol does not rely on GHZ states for security. Alice then computes the parameter $r$ as
    \begin{equation}
    r = (B_{\text{decimal}} \times K) + t_{\text{hide}},
    \end{equation}
    where $B_{\text{decimal}}$  is $B$ in decimal representation. Alice chooses $\beta$ as in Eq. \ref{beta}
and applies the unitary
    \begin{equation}
        \ket{\psi_{t_{\text{hide}}}} = U_{\text{QKT}}(\beta)^{t_{\text{hide}}} \ket{\psi_0},
    \end{equation} 
entangling the qubits in the process. Further, the initial state is effectively scrambled by this unitary operator concealing the original message. Hence, we shall call $U_{\text{QKT}}(\beta)^{t_{\text{hide}}}$ the concealment unitary. Now, the evolved state $\ket{\psi_{t_{\text{hide}}}}$ holds the encrypted version of the original message $\mathcal{I}$. This state can now be stored in a quantum memory or transmitted to Bob for further storage or decryption.

\begin{figure*}[t]
    \centering
\includegraphics[width=\linewidth]{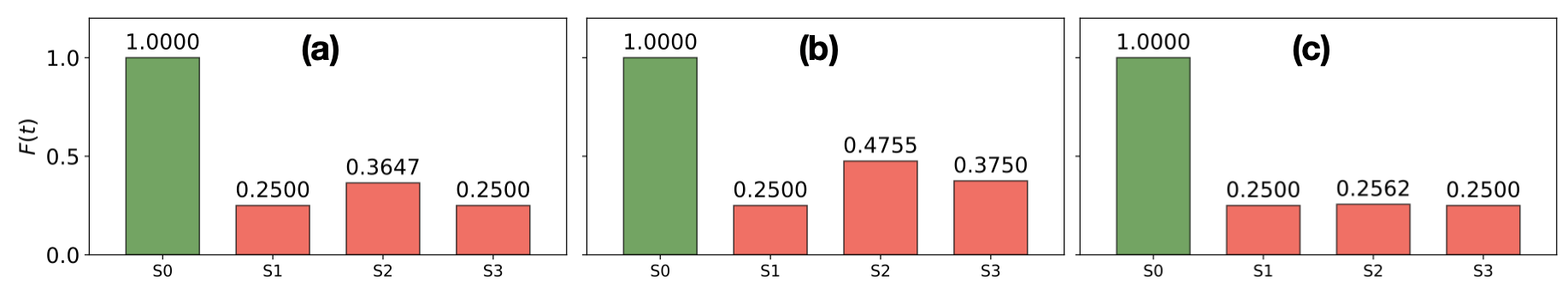}
    \caption{Security of the four qubit encryption corresponding to the parameters in Fig. \ref{fig:exact_recorrence}.  Fidelity of recovered state is shown for $4$ different scenarios.
    S0 (green bar) represents message recovery by Bob who has access to all the keys. Red bars correspond to three representative attacks by Eve based on incorrect choice of parameters: (S1) incorrect recovery time $T-t_\text{hide}$, (S2) incorrect scrambling parameter $\beta$, and (S3) direct measurement without knowing any parameters. In S1-S3, Eve cannot recover the message with sufficiently high fidelity. }
    \label{fig:security1}
\end{figure*}

\subsection{Decryption through quantum recurrence}   
{Alice transmits the state to Bob via a quantum channel.}
Bob also receives $r$  from Alice in person or via a classical channel, and already knows the PSK $K$ and $j$.  Using these, Bob computes the parameters
\begin{equation}
\beta = 4\pi  j\frac{  r}{s}, \;\;\; \text{and} \;\;\; t_{\rm{hide}} = r \mod K    
\end{equation}
required to apply the unitary operator on the state $\ket{\psi_{t_{\text{hide}}}}$ he received from Alice. Bob then applies the recovery unitary until the near-recurrence time to retrieve
    \begin{equation}
    |\Psi_{\text{rec}}\rangle = U_{\text{QKT}}(\beta)^{T - t_{\text{hide}}} ~ |\Psi(t_{\text{hide}})\rangle.
    \end{equation}
Assuming that the state that Bob received was not tampered or decohered, then $|\Psi_{\text{rec}}\rangle \approx \ket{\psi_0}$.  Next, Bob will be able to extract the message by inferring the qubit positions where actual message was stored, i.e, by computing the bitmask to be
\begin{equation}
 B_{\text{decimal}} = \lfloor r / K \rfloor.   
\end{equation}
He measures only the qubits specified by bitmask $B$ in the computational basis to recover the message contained in $|\psi_0\rangle$ {as shown in Fig. \ref{fig:almost_periodic}}

\section{Security of the protocol}
Of the $N$ qubits used for the initial state, the message is embedded in $m$ qubits  (see Eq. \ref{eq:initial_state}). Since Eve does not know the correct $m$ qubits which contain the message $\mathcal{I}_b$, she cannot recover the encrypted message even if she somehow learns the values of $r$ and $T$. During the transmission of the message, the security of the message is guaranteed by the nature of the unitary operator. Note that the encrypted state $\ket{\psi_{t_{\text{hide}}}}$ is entangled by the concealment unitary operator. Anyone accessing the encrypted state during transmission will still not be able to access the message since any subsystem projections of the entangled state would give mixed states, effectively giving out no information. Moreover, any tampering by Eve will be detectable by Bob. Since Bob now knows the location of ancilla qubits, he can measure them individually, as $\ket{\psi_0}$ is a product state, and verify that they are  $\ket{0}$ as initialized by Alice. Computing the fidelity of known test states is another possibility.

Figure \ref{fig:security1} demonstrates the security offered by the QRE scheme in the case when exact recurrence occurs due to quantum resonance in the kicked top. In this case, the initial state is guaranteed to recur with an exact periodicity of $T$. Eve's attempts to intrude would always fail without the knowledge of {\it all} the parameters and how the information is actually encoded in the initial state. Three possibilities are shown in Fig. \ref{fig:security1} corresponding to the parameters $t_{\rm hide}, \beta$ and location of actual message in initial state indicated by bitmask $B$. Consider the following scenarios: (S1) Eve guesses all the parameters correctly except for  $t_{\rm hide}$, (S2) Eve guesses all the parameters correctly except for $\beta$, and (S3) Eve does not know any parameters, and performs a direct measurement of the quantum state transmitted between Alice and Bob, trying to retrieve information. As Fig. \ref{fig:security1} shows, under none of these circumstances, Eve can decrypt the message with fidelity exceeding about 0.4. This implies that she will end up with a very poor replica of the initial state that is not useful in practice to infer the secret message.

\begin{figure*}[t]
    \centering
\includegraphics[width=0.97\linewidth]{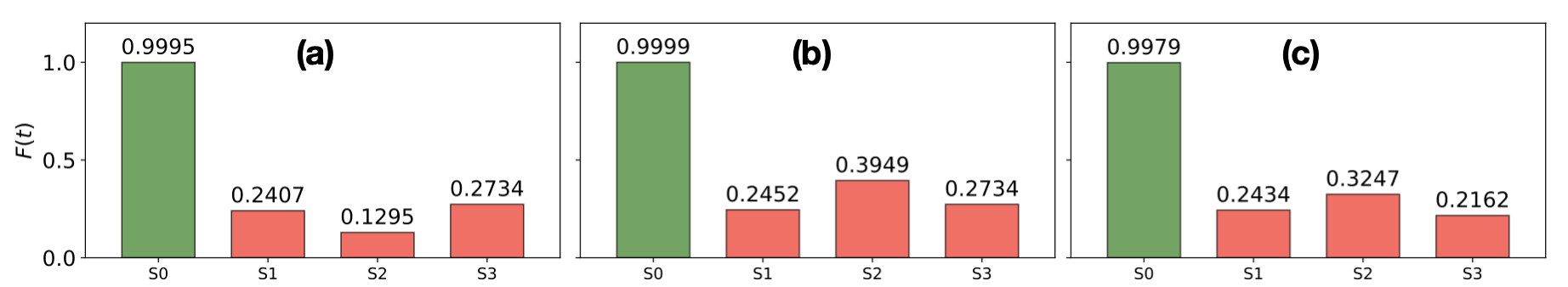}
    \caption{Security of the four qubit encryption corresponding to the parameters in Fig. \ref{fig:almost_periodic}.  Fidelity of recovered state is shown for $4$ different scenarios.
    S0 (green bar) represents message recovery by Bob who has access to all the keys. Red bars correspond to three representative attacks by Eve based on incorrect choice of parameters: (S1) incorrect recovery time $T-t_\text{hide}$, (S2) incorrect scrambling parameter $\beta$, and (S3) direct measurement without knowing any parameters. In S1-S3, Eve cannot recover the message with sufficiently high fidelity. }
    \label{fig:security2}
\end{figure*}

Similarly, the security offered by the protocol can be demonstrated when the initial state displays near-recurrence after some time $T$. Since the recurrence time is state-dependent, Alice has to also send  $T$ along with $r$, for Bob to recover the message. In this case, the fidelity reaches quite close to unity after time $T$. Near-recurrence of the initial state $|\psi_0\rangle$ with fidelity  close to unity is sufficient for Bob to recover the encrypted state.
For same reasons as before, not possessing all the keys, Eve will not be able to decrypt the message. As shown in Fig. \ref{fig:security2}, in all the scenarios of S1, S2, and S3, Eve will not be able to attain fidelity greater than 0.3. This is insufficient to decrypt the message.

\section{Quantum teleportation of encrypted state} 
\label{sec: 4}
\begin{figure}[t]
    \centering    
    \includegraphics[width=0.9\linewidth]{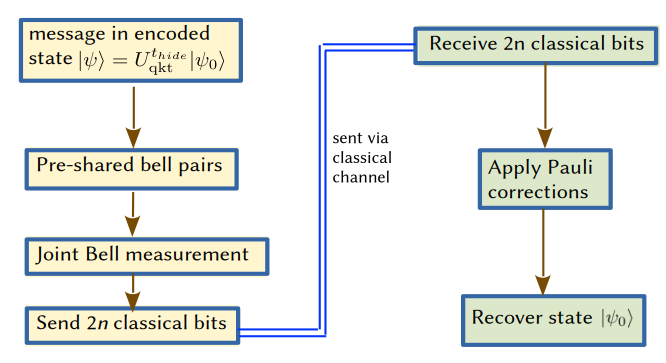}
    \caption{Communication protocol with quantum resonance encryption. Yellow boxes are tasks done by Alice, and green boxes are tasks done by Bob. }
    \label{fig: teleportation}
\end{figure}

Once QRE has been applied to an initial state $\psi_0\rangle$, Alice can use quantum teleportation protocol \cite{PhysRevLett.70.1895} to send encrypted state $\ket{\psi_{t_{\text{hide}}}}$ to Bob. To do this, Alice and Bob have to pre-share Bell pairs equal to the length of the message (including the ancilla qubits). 
To send a single bit message to Bob, Alice first encodes the classical bit as before. For ease of demonstration, suppose that she adds only one ancilla qubit for increasing the Hilbert space  dimension to securely conceal the message qubit. She then applies  $U_{\text{QKT}}(\beta)^{t_{\text{hide}}}$ to entangle the qubits, where $\beta$ is chosen as in Eq. \ref{beta}.
The 2-qubit entangled state after $t_\textrm{hide}$ steps that Alice wants to teleport has the following form:
\begin{equation}
\ket{\psi}_{\text{message}} = \alpha\ket{00}_{M_1M_2} + \beta\ket{11}_{M_1M_2},
\end{equation}
where $|\alpha|^2 + |\beta|^2 = 1$. In the following, for simplicity, we demonstrate this particular instance of a two-qubit communication protocol. This follows the series teleportation framework \cite{lee2000entanglement,lee2002multipartite}, a direct multi-qubit generalization of the standard single qubit teleportation protocol \cite{bennett1993teleporting}.  It is shown as a schematic in Fig. \ref{fig: teleportation}.  

In this protocol, Alice and Bob share two pre-established Bell pairs:
\begin{align}
   \ket{\Phi^+}_{A_1B_1} &= \frac{1}{\sqrt{2}}(\ket{00}_{A_1B_1} + \ket{11}_{A_1B_1}) \nonumber\\
   \ket{\Phi^+}_{A_2B_2} &= \frac{1}{\sqrt{2}}(\ket{00}_{A_2B_2} + \ket{11}_{A_2B_2}).
\end{align}
The complete initial state is:
\begin{equation}
    \ket{\Psi_0} = \ket{\psi}_{M_1M_2} \otimes \ket{\Phi^+}_{A_1B_1} \otimes \ket{\Phi^+}_{A_2B_2}. 
\end{equation}
Alice now performs a {joint measurement} projecting the two pairs $(M_1, A_1)$ and $(M_2, A_2)$ onto the Bell basis $\{\ket{\Phi^+},\ket{\Phi^-}, \ket{\Psi^+}, \ket{\Psi^-}\}$.
Here, 
$
    \ket{\Phi^\pm} = \frac{1}{\sqrt{2}}(\ket{00} \pm \ket{11} ), \:\:
     \ket{\Psi^\pm} = \frac{1}{\sqrt{2}}(\ket{01} \pm \ket{10}).
$
The measurement has $4 \times 4 = 16$ possible outcomes (one Bell state for each pair).
After projection, the complete state becomes
\begin{equation}
    \ket{\Psi} \xrightarrow{\text{Bell measurement}} \ket{\text{Bell}_1}_{M_1A_1} \otimes \ket{\text{Bell}_2}_{M_2A_2} \otimes \ket{B_1B_2},
\end{equation}
where $\text{Bell}_1, \text{Bell}_2 \in \{\Phi^+, \Phi^-, \Psi^+, \Psi^-\}$. Correspondingly, Alice gets 4 classical bits, where 2 bits encode the
 Bell state for $(M_1, A_1)$, and the remaining two bits encode the Bell state for $(M_2, A_2).$
For {any} measurement outcome, Bob's qubits collapse to
\begin{equation}
    \ket{B_1B_2} = \sigma_1 \otimes \sigma_2 (\alpha\ket{00}_{B_1B_2} + \beta\ket{11}_{B_1B_2}),
\end{equation}
where $\sigma_1, \sigma_2 \in \{I, X, Y, Z\}$ are Pauli operators determined by the measurement outcome. Bob now applies the inverse Pauli operations based on the classical communication from Alice:
\begin{equation}
\ket{B_1B_2}^{\text{recovered}} = \sigma_1^\dagger \otimes \sigma_2^\dagger \ket{B_1B_2} = \alpha\ket{00}_{B_1B_2} + \beta\ket{11}_{B_1B_2}. \nonumber
\end{equation} 
The Table \ref{tab:table1} shows the operations to be done by Bob depending on the classical bits received from Alice. Bob then completes the remaining evolution by doing $U_{\text{QKT}}(\beta)^{T - t_{\text{hide}}}\ket{B_1B_2}^{\text{recovered}}$ to get back the unentangled two qubit state from Alice. Since Bob knows which of the two qubits  carry the message using $\beta$ and the pre-shared parameters, he can perfectly recover the message.

\begin{table}[b]
\centering
\begin{tabular}{c|c|c}
\hline
Bell state for $(M_i, A_i)$ & Alice's bits  & Bob's correction $\sigma_i$ \\
\hline
$\ket{\Phi^+}$ & $(0,0)$ & $I$ \\
$\ket{\Phi^-}$ & $(1,0)$ & $Z$ \\
$\ket{\Psi^+}$ & $(0,1)$ & $X$ \\
$\ket{\Psi^-}$ & $(1,1)$ & $XZ$ (or $iY$)
\end{tabular} 
\caption{Operations to be done by Bob on his qubits depending on the classical bits received from Alice.}
\label{tab:table1}
\end{table}

 \section{Quantum key distribution (QKD) using QRE} 
 \label{sec: 5}  
 QKD protocols generate a completely random, secure classical key shared between two parties, which can be used to encrypt classical messages.  In the seminal BB84 protocol \cite{bennett2014quantum}, the classical key is established by preparing and measuring random qubits across two random bases, announcing the basis over a classical channel, retaining the outcomes where the bases matched, and comparing a subset of those outcomes to estimate the error rate \cite{nielsen2010quantum, BENNETT20147}.
 
 QRE can naturally act as a QKD generator. For this, Alice prepares a random $m-$bit message, and encodes onto a $N-$qubit quantum register, along with $N-m$ ancilla qubits at random positions, initialized to $\ket{0}$ as before. She then sends the scrambled state to Bob, who recovers the initial state using the resonance condition and then measures all the qubits in the computational basis. The resulting outcome can be used as $m-$bit random key, corresponding to the random message bits.  The aniclla bits must all measure $\ket{0}$, and any deviation indicates the presence of Eve. Classical communication required to establish the key is minimal in QRE compared to BB84, as only the $\beta$ value needs to be sent to Bob. QRE also does not require comparing a subset of measurements to detect an eavesdropper, as resonance failure has a built-in intrinsic mechanism to identify tampering.

\section{Robustness of the protocol} 
 \label{sec: 6} 
 While the quantum resonance based encryption is theoretically sound, practical considerations require that the protocol be robust to small errors or fluctuations in $\beta$ that are unavoidable in an experimental setting. It is important to note that the robustness is not guaranteed for all the pairs of $(r,s)$ constituting the resonance condition. Furthermore, the near-recurent timescale  though finite, can be large for arbitrary states and tolerance $\epsilon$. As the pre-shared integer $K$ must satisfy the condition $K>T$ for obtaining the correct bitmask, the protocol requires that it must be fixed independently of the transmitted message. The protocol requires prior knowledge of an upper bound on the admissible recurrence times $T$. Since both these factors are important from an experimental perspective, we shall focus on a subset of $(r,s)$ that give exact recurrence of the kicked top Floquet operator, with a state-independent recurrence time that is known {\it apriori}.
 
Exact recurrence has been identified only for a few values of the kicking strength so far \cite{anand2024quantum}, namely at $\beta= 2\pi J, \: \pi J $ and $\pi J/2$.  At these $\beta$ values, $U_\mathrm{QKT}^T=I$ for a finite $T$, independent of the initial state. All Floquet eigenphases are commensurate, i.e., $\phi_k = 2\pi n_k / T$
for integers $n_k$. Under a small detuning $\beta \rightarrow \beta+\delta \beta$, the eigenphases acquire only small shifts, so the operator $U_\mathrm{QKT}^T$ remains close to the identity and the recurrence fidelity decays smoothly. Exact recurrences can also occur at other $\beta$ values.   Since $\beta= 4\pi J\frac{r}{s},$ the Floquet operator depends only on the equivalence class of $r \: (\mathrm{mod}\: s)$. Consequently, many apparently distinct resonance parameters generate the same finite-order Floquet dynamics and belong to the same equivalence class. Hence, the additional recurrence points represent alternative parametrizations of a small number of underlying finite-order Floquet operators rather than genuinely new exact-recurrences. For instance, in Fig. \ref{fig:exact_recorrence}(c), $\beta= 9\pi/2= 3\pi + \frac{3\pi}{2}$, which is the same as $\pi J/2$, for $j=3,$ upto an integer multiple of $\pi.$ 

\begin{figure}[t]
    \centering
    \includegraphics[width=0.9\linewidth]{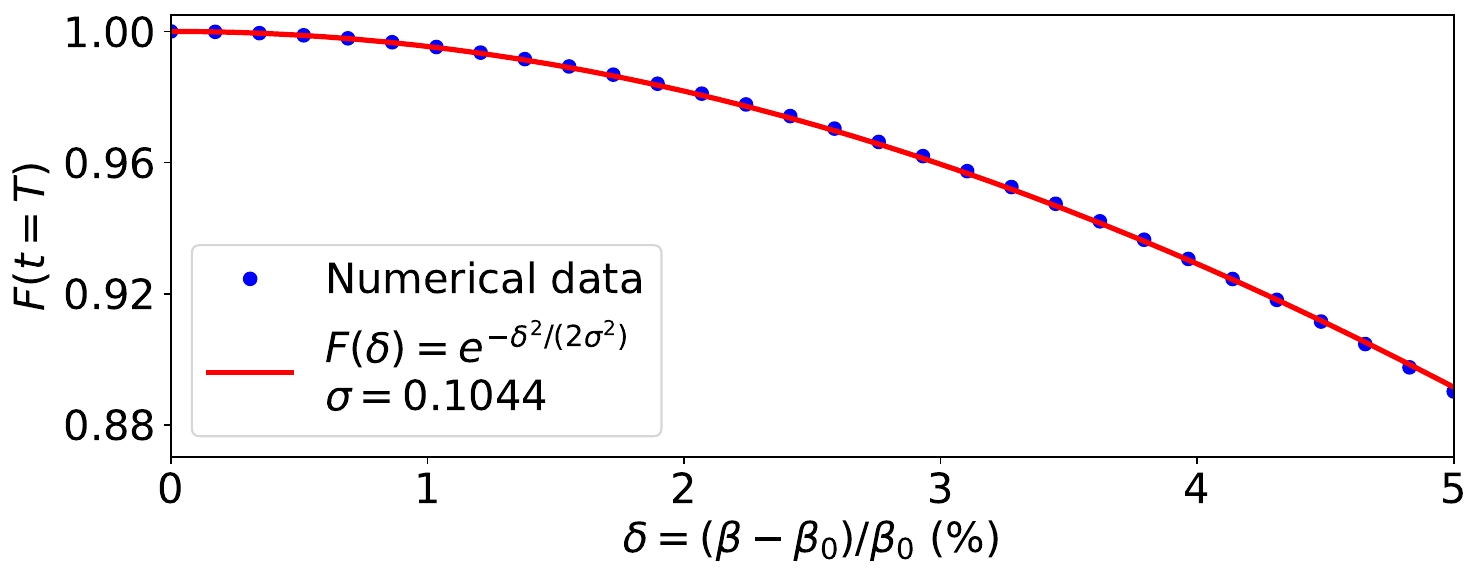}
    \caption{Decay of the fidelity of recurrence (at $t=T$) under perturbation at $\beta= \pi J/2,$ with $J=2$. Fidelity $\sim 0.9$ is achieved with a $5\%$ error in $\beta$ shows the robustness of the protocol.}
    \label{fig:robustness}
\end{figure}

Robustness at exact recurrence is shown in Fig. \ref{fig:robustness}, 
which displays the effect of deviation from correct value of $\beta$, denoted as
$\delta = (\beta-\beta_0)/\beta_0$. Thus, exact recurrence occurs at $\delta=0$ when $\beta= \pi J/2,$ with $j=2$. The figure shows that the fidelity decays with increase in $\delta$, and decay is well approximated by a one-sided Gaussian of the form $F(\delta) \sim e^{-\delta^2/(2 \sigma^2)}$, where $\sigma$ is the standard deviation.  As this figure shows, following a Gaussian decay, a recurrence fidelity of $F\approx0.9$ can be achieved with $5\%$ error in the kick-strength. Since the relative calibration tolerance scales as $
\frac{\Delta\beta}{\beta}
= \frac{s\Delta \beta}{4 \pi J r},$ Alice should use $r_{\mathrm{eff}}= r (\mathrm{mod} \:s)$ to construct $\beta,$ as $r$ and $r_\mathrm{eff}$ have the same dynamics.  She can still send the inflated $r$ over the classical channel, as this protects Eve from immediately discerning the equivalence class of $\beta.$ Bob, knowing the pre-shared $s$, can find out $r_\mathrm{eff}$ for the recovery. However, since  there are only three equivalence classes identified so far, Eve may guess the correct dynamics with $1/3$ probability. But successful message recovery additionally requires determination of the hidden evolution time and bit-mask information. Consequently, knowledge of the recurrence class alone is insufficient to reconstruct the encoded message.

In the exact recurrence scenario,  since the  time period is state independent, and purely a function of the dynamics,  Bob can infer the time period from $\beta.$ Therefore, Alice only has to transmit $r,$ and consequently, the classical overhead is lower compared to the general resonance protocol. Thus, while the general resonances provide a larger space of admissible scrambling parameters, making the effective dynamics less predictable from intercepted classical information, the exact recurrence dynamics wins  in experimental robustness, bounded time period,  and classical overhead.

 \begin{comment}
 \textcolor{blue}{At other resonant values of $\beta$ determined by $(r,s)$ pairs that do not yield a finite-order Floquet operator, the revival time is state dependent. Very fidelity revivals with $F(T) \approx 1$ are obtained when  $\bra{\psi_0}U_{\mathrm{QKT}}^T\ket{\psi_0} \approx 1.$ This recovery is not stemming from a stronger $U^T_\mathrm{QKT}=1$ condition. Consequently, different initial states may exhibit different recurrence times. While the state dependent recurrence time makes the encryption stronger from a cryptography perspective, they are generally more sensitive to perturbations, since these revivals are not associated with finite-order recurrences of the Floquet operator.}
 \end{comment}
 \subsection{ Experimental outlook}
{The data storage and direct communication protocols using QRE  assumes hardware such as quantum memory and long-distance teleportation of multiple quboits,  which are in an early stage of development. A working quantum memory is fundamental to ensure data security in the post quantum era, and also to realize theoretically promised efficiency and speedup from quantum computers. Proof-of-principle experiments demonstrating quantum memory is already available \cite{bozkurt2025mechanical,mariantoni2011implementing,yao2022experimental,freer2017single}.}
 
{ Similarly, significant progress in experimental implementations are presently being made in long-distance teleportation \cite{pirandola2015advances}, from fiber networks \cite{valivarthi2016quantum} that extend to about a hundred kilometers to free-space communications between earth and a satellite \cite{ren2017ground}, spanning over a thousand kilometers.   The main challenge in achieving long-distance teleportation are photon loss in optical fibers and beam divergence in free space \cite{weston2018heralded, ren2017ground}. In the case of multi-qubit teleportation, fidelity loss due to decoherence is a huge issue, as one has to store the qubits that may arrive at different times and use them together in a functional application. Experimentally  achieved multi-qubit teleportation so far include a two-qubit teleportation using a single photon, where multiple degrees of freedom (spin and orbital angular momentum) are simultaneously teleported \cite{wang2015quantum}, multi-qubit teleportation using cluster states \cite{kumar2020experimental}, and controlled quantum teleportation (CQT) using multipartite entanglement of GHZ states \cite{yonezawa2004demonstration,barasinski2019demonstration}. Rapid improvements in the technology can be expected in the next decade.} The QRE protocol is amenable to experimental validation using currently available quantum platforms, while larger-scale implementations can be pursued as quantum control and communication technologies mature.

\section{Conclusion} 
The quantum resonance encryption discussed in this letter relies on quantum information scrambling and quantum resonance to provide layers of security. This is in addition to benefiting from the no-cloning theorem (Eve cannot copy the scrambled state), and measurement disturbance (Eve's tampering can be detected), which are characteristics of quantum key distribution protocols. The classical communication overhead (key length) in QRE is small and independent of message length, unlike the quantum one-time pad (QOTP), an encryption protocol \cite{boykin2003optimal} considered informationally secure. In QOTP,  the length of the key scales as twice the number of qubits in the message  \cite{boykin2003optimal}. There is also a quantum direct communication version of QOTP \cite{deng2004secure}, without a classical key. The security check to authenticate the channel, however, is mainly done at the pre-encoding stage, and it can fail if Eve waits to attack until the actual message is sent. Also, in this protocol \cite {deng2004secure}, the message qubits travel back and forth via the quantum channel, reducing the effective communication range due to additional photon loss and decoherence. QRE works both as an encryption and a direct communication scheme (with a small, message-independent key), and it does not have the above drawbacks. Moreover, it can also be used for quantum key distribution.

%\textcolor{blue}{  Quantum resonance encryption utilizes the dynamics of a physical system, and conceals it within the vastness of an extended Hilbert space. It can be naturally recovered via the resonance dynamics only by an authorized person. The scheme includes an in-built tampering detection using the periodicity of the dynamics.}

{The quantum kicked top model, the dynamical system used, is experimentally amenable across various quantum computing platforms \cite{chaudhury2009quantum,neill2016ergodic,krithika2019nmr}, which enhances the practicability of the protocol. Each of the other hardware requirements in the proposal has been separately demonstrated experimentally as proof-of-principle realizations. Furthermore, quantum  memory and quantum networks assumed in the protocols are essential components of the future quantum hardware, and are intensely pursued.
%This work is crucial for proactive post-quantum preparedness.
%Assuming hardware components that do not currently exist at a practical scale are justified, as we need to ensure privacy and data security from the outset of the next generation of quantum computers.}

The present work focuses on the dynamical mechanism of information concealment and recovery based on quantum resonance.  While the protocol assesses security though recovery fidelity and tamper detection, a full information-theoretic security analysis against arbitrary coherent attacks remains an interesting direction for future work.
\FloatBarrier
\section{Acknowledgment} Authors acknowledge the I-HUB quantum technology foundation (I-HUB QTF) based in IISER Pune for financial support.
\bibliography{ref}
\bibliographystyle{unsrt}
\end{document}